\documentclass[aps,prl,reprint,twocolumn,showpacs,superscriptaddress,amsmath,amssymb,nofootinbib]{revtex4-2}

\usepackage{graphicx}
\usepackage{dcolumn}
\usepackage{bm}

\usepackage{hyperref}
\usepackage{lineno}  
\usepackage{xfrac}
\usepackage{soul}
\usepackage{xcolor}
\usepackage{amsfonts}
\usepackage{verbatim}
\usepackage{graphicx}

\newcommand{\strf}{\mathrm}

\DeclareMathOperator{\tr}{tr}

\newcommand{\detailtexcount}[1]{%
  \immediate\write18{texcount -merge -sum -q manuscript-v5-prl-v2.tex output.bbl > manuscript-v5-prl-v2.wcdetail }%
  \verbatiminput{manuscript-v5-prl-v2.wcdetail}%
}

\newcommand{\quickwordcount}[1]{%
  \immediate\write18{texcount -1 -sum -merge -q manuscript-v5-prl-v2.tex output.bbl > manuscript-v5-prl-v2-words.sum }%
  \input{manuscript-v5-prl-v2-words.sum} words%
}

\newcommand{\quickcharcount}[1]{%
  \immediate\write18{texcount -1 -sum -merge -char -q manuscript-v5-prl-v2.tex output.bbl > manuscript-v5-prl-v2-chars.sum }%
  \input{manuscript-v5-prl-v2-chars.sum} characters (not including spaces)%
}

\usepackage{ifthen}
\let\oldtextcolor\textcolor 
\renewcommand{\textcolor}[2]{
    \ifthenelse{\equal{red}{#1}}
    {\suppress#2\endsuppress}
    {\oldtextcolor{#1}{#2}}%
}
\makeatletter   
\font\dummyft@=dummy \relax
\def\suppress{%
    \begingroup\par
    \parskip\z@
    \offinterlineskip
    \baselineskip=\z@skip
    \lineskip=\z@skip
    \lineskiplimit=\maxdimen
    \dummyft@
    \count@\sixt@@n
    \loop\ifnum\count@ >\z@
    \advance\count@\m@ne
    \textfont\count@\dummyft@
    \scriptfont\count@\dummyft@
    \scriptscriptfont\count@\dummyft@
    \repeat
    \let\selectfont\relax
    \let\mathversion\@gobble
    \let\getanddefine@fonts\@gobbletwo
    \tracinglostchars\z@
    \frenchspacing
    \hbadness\@M}
\def\endsuppress{\par\endgroup}
\makeatother

\begin{document}

\title{{\it Fibrotaxis}: gradient-free, spontaneous and controllable droplet motion on soft solids}

\author{S. R. Bhopalam}
\affiliation{School of Mechanical Engineering, Purdue University, West Lafayette, IN 47907, USA.}
\author{J. Bueno}
\affiliation{Midge Medical GmbH, Colditzstra{\ss}e 34/36, 16A, Berlin 12099, Germany.}
\author{H. Gomez} \thanks{Corresponding author}
\email{hectorgomez@purdue.edu}
\affiliation{School of Mechanical Engineering, Purdue University, West Lafayette, IN 47907, USA.}

\date{\today}

\begin{abstract}
{Most passive droplet transport strategies rely on spatial variations of material properties to drive droplet motion, leading to gradient-based mechanisms with intrinsic length scales that limit the droplet velocity or the transport distance. Here, we propose droplet {\it fibrotaxis}, a novel mechanism that leverages an anisotropic fiber-reinforced deformable solid to achieve spontaneous and gradient-free droplet transport. Using high-fidelity simulations, we {identify the fluid wettability, fiber orientation, anisotropy strength and elastocapillary number as critical parameters} that enable controllable droplet velocity and long-range droplet transport. Our results highlight the potential of fibrotaxis as a droplet transport mechanism that can have a strong impact on self-cleaning surfaces, water harvesting and medical diagnostics.}
\end{abstract}

\maketitle

\section*{Introduction}

The control and movement of small droplets is critical in many engineering and science applications, including microfluidics \cite{joanicot_microfluidics_2005, stone_arfm_2004, seemann_etal_repphys_2011}, microfabrication \cite{srinivasrao_etal_sci_2001}, medical diagnostics \cite{serra2017power}, and drug delivery \cite{li_etal_comphys_2018}. While controllable droplet manipulation on solid surfaces has been achieved using, e.g., gradients of material properties or external force fields, a gradient-free mechanism that passively induces controllable droplet motion would open opportunities for transformative advances. Here, we propose droplet fibrotaxis---a novel, gradient-free transport mechanism for small droplets that enables droplet motion with controllable direction and speed without relying on external force fields.

Over the last few years, notable advances in droplet manipulation have been achieved using electric fields \cite{li_etal_electric_nat_2019}, light \cite{li_etal_sciadv_light_2020}, vibrations \cite{john_thiele_vibrat_prl_2010, brunet_etal_prl_2007}, as well as gradients of curvature \cite{cunjing_etal_prl_2014}, topography \cite{li_etal_topography_natphys_2016}, confinement \cite{dangla_etal_pnas_2013}, wettability \cite{daniel_wetgradtemp_sci_2001, thiele_etal_prl_2004}, and surface charge density \cite{sun_etal_natmat_2019}. A particularly interesting example is droplet durotaxis \cite{style_etal_2013}, which is a passive droplet transport mechanism that uses stiffness gradients of the solid substrate. Durotactic droplet motion relies on the ability of the droplet to deform the solid through elastocapillary interactions that emerge from forces at the fluid-fluid interface \cite{bico2018elastocapillarity}. Experimental \cite{style_etal_2013} and computational \cite{bueno_2018a} studies have shown that small droplets placed on a soft silicone gel with a stiffness gradient move spontaneously without application of external forces, traveling up or down the gradient depending on the droplet's wettability. More recent research has proposed new droplet motion mechanisms based on elastocapillarity, including systems based on gradients of strain (tensotaxis) \cite{bueno_etal_2017}, gradients of pressure (bendotaxis) \cite{bradley_2019} or substrate stretching \cite{smith_prl_2021}.

Fibrotaxis, the droplet transport mechanism proposed herein, emerges from elastocapillary interactions with soft, homogeneous, anisotropic solids. One way to understand soft anisotropic solids is by conceptualizing them as an isotropic matrix with an infinite number of fibers continuously distributed. Fibrotaxis draws inspiration from cell migration in an extracellular matrix, a highly anisotropic network of collagen \cite{saez_pnas_2007, sopher_etal_biophys_2018}, but spontaneous droplet motion on soft anisotropic solids has not been explored. Our data show that droplet fibrotaxis features unique advantages, such as, independence of external forces, controllable droplet direction and regulable velocity. More importantly, while gradient-based mechanisms are limited by inherent length scales that restrict the maximum transport distance or velocity, fibrotaxis is a gradient-free transport process. To elucidate the mechanism underpinning fibrotaxis we develop a high-fidelity fluid-structure interaction model that accounts for the coupled dynamics of a nonlinear, anisotropic hyperelastic solid, and two immiscible fluids with surface tension at their interface. Our results for droplets moving on fiber-reinforced gels unveil a complex interaction between the droplet's wettability and the orientation of the fibers that permits to control the droplet direction and velocity. Together, our findings indicate that fibrotaxis is a robust mechanism for droplet transport with unique features that can open new opportunities in microfluidics, medical diagnostics and energy generation.

\section{Model of droplet fibrotaxis}

\subsection*{Elastocapillary wetting}

\begin{figure}
    \includegraphics[width=0.95\linewidth]{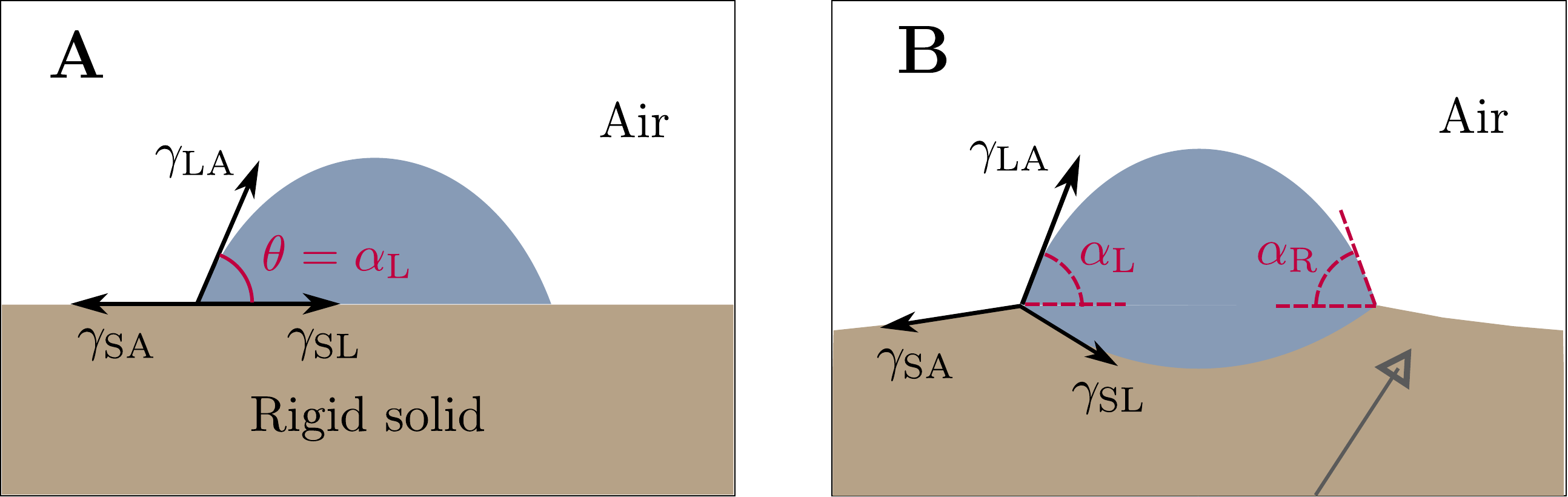}
    \caption{Schematic of a droplet wetting on (A) rigid solid and a (B) fiber-reinforced deformable solid. {In (B), gray colored arrow depicts the orientation of fibers. This depiction does not indicate that we model the fibers discretely; rather, the model assumes that there are infinite fibers continuously distributed}.}
    \label{fig:schematic}
\end{figure}

A liquid droplet resting on a flat and rigid solid surrounded by air will form a contact angle at equilibrium that satisfies the Young-Dupr\'e equation $\gamma_{\mathrm{SL}} + \gamma_{\mathrm{LA}}\cos\theta = \gamma_{\mathrm{SA}}$. Here, $\theta$ is the static contact angle between the liquid-air interface and the solid, while $\gamma_{\mathrm{SL}}$, $\gamma_{\mathrm{LA}}$ and $\gamma_{\mathrm{SA}}$ are the surface tensions at the solid-liquid, liquid-air, and solid-air interfaces, respectively; see Fig.~\ref{fig:schematic}A. When the solid is deformable, however, the process is controlled by elastocapillary wetting \cite{style_arfm_review_2017}, and the Young-Dupr\'e equation breaks down. Capillary forces at the fluid-fluid interface pull up the solid at the contact line, creating a ridge. Simultaneously, the overpressure in the droplet's interior generates a dimple in the solid located under the droplet. The combined effect of these two actions produces a rotation of the interfaces near the contact line that gives rise to an apparent contact angle $\alpha$ that differs from $\theta$; see Fig.~\ref{fig:schematic}B. We hypothesize that when a droplet is placed on a soft anisotropic solid, such as a fiber-reinforced polymer, the apparent contact angles on the left ($\alpha_{\mathrm{L}}$) and right ($\alpha_{\mathrm{R}}$) triple points differ, leading to an imbalance of the horizontal forces that produces droplet motion. To test this hypothesis, we propose a three-dimensional multi-component fluid-solid interaction model. We consider a system comprising of two immiscible fluids, i.e., liquid droplet and air, in contact with a solid. We use the phase-field method \cite{anderson_arfm_1998} to model the fluids, by replacing the liquid-air interface with a thin transition region. We model the fiber-reinforced anisotropic solid using a transversely isotropic hyperelastic material.

\subsection{Governing equations of fluid mechanics}

We describe the dynamics of the droplet and air using the Navier-Stokes-Cahn-Hilliard (NSCH) equations, a phase-field model for flow of two immiscible fluids \cite{jacqmin_jcp_1999}. The NSCH equations assume that the two fluids are incompressible and share the same velocity field. The governing equations in Eulerian coordinates are 
\begin{subequations}
    \begin{align}
        &\nabla \cdot \bm{v} = 0, 
        \label{eqn:nsch_divfree} \\
        &\rho \Big(\partial_t \bm{v} + \bm{v} \cdot \nabla \bm{v}\Big)  = \nabla \cdot \bm{\sigma}^f+\rho\bm{g}, 
        \label{eqn:nsch_momeqn} \\ 
        &\begin{aligned}
            \partial_t c + \nabla\cdot(\bm{v} c) & = M\gamma_{\mathrm{LA}} \Delta \left(-\frac{3}{2}\epsilon\Delta c \right) \\
            & + M\gamma_{\mathrm{LA}} \Delta \left(\frac{24}{\epsilon} c(1 - c)(1 - 2c) \right). 
        \end{aligned}
        \label{eqn:nsch_phfield}
    \end{align}
    \label{eqn:nsch}
\end{subequations}
$\!\!\!$Here, $\bm{v}$ is the fluid velocity; $\rho$ is the fluid density; $\partial_t$ denotes partial differentiation with respect to time; $\bm{\sigma}^f$ is the fluid Cauchy stress tensor; $\bm{g}$ is the acceleration of gravity; $c \in [0,1]$ is the phase-field denoting the volume fraction of liquid; $\epsilon$ is the diffuse interface length scale; and $M$ is the mobility coefficient. The fluid Cauchy stress tensor is defined as $\bm{\sigma}^f = - p \bm{I} + 2\eta \nabla^{s} \bm{v} - \frac{3}{2}\gamma_{\mathrm{LA}}\epsilon \nabla c \otimes \nabla c$, where $p$ is the pressure and $\nabla^{s}$ is the symmetrization of the spatial gradient operator, $\eta$ is the dynamic viscosity of the mixture given by $\eta = \eta_1 c + \eta_2 (1 - c)$, where $\eta_1$ and $\eta_2$ are the dynamic viscosities of the liquid and air, respectively. Because all of our simulations are performed at length scales $\ell<1$ mm, gravity forces are negligible and we will ignore them in our calculations.

\subsection{Governing equations of solid mechanics}

We describe the dynamics of the solid using the linear momentum balance equation in Lagrangian coordinates,
\begin{equation}
    \rho_0^s \left. \partial_t^2 \bm{u} \right|_{\bm{X}} = \nabla_{\bm{X}} \cdot \bm{P},
    \label{eqn:solidmech_linearmom}
\end{equation}
\noindent where $\rho_0^s$ is the mass density of the solid in the undeformed configuration, $\bm{u}$ is the solid displacement and $\bm{P}$ is the first Piola-Kirchhoff stress tensor. In Eq.~\eqref{eqn:solidmech_linearmom}, we use the spatial derivative with respect to the material coordinates $\bm{X}$ in $\nabla_{\bm{X}}$. The subscript $\bm{X}$ in $\left. \partial_t^2 \bm{u} \right|_{\bm{X}}$ indicates that the second-order time derivative is taken by holding $\bm{X}$ fixed.  

We model the solid as an anisotropic material. Our model is based on the theory of fiber-reinforced composites \cite{holzapfel_etal_jelast_2000, gasser_etal_jrsc_2006} and has been widely used for anisotropic gels. We consider a material composed of an isotropic matrix, in which one or several families of fibers are homogeneously distributed along specific directions. We do not model the fibers discretely; rather, the model assumes that there are infinite fibers continuously distributed. Here, we use a single family of fibers (i.e., all fibers have the same orientation) to represent a transversely isotropic material \cite{nolan_2014}, the simplest form of an anisotropic material. The fibers in the undeformed configuration are oriented along the direction defined by the unit vector $\bm{d}$. We define the orientation tensor $\bm{A} = \bm{d} \otimes \bm{d}$. The strain energy density of the solid is given by \cite{nolan_2014}
\begin{equation}
    W = W_{\mathrm{vol}} + W_{\mathrm{iso}} + W_{\mathrm{aniso}},
    \label{eqn:w_solidgeneral}
\end{equation}
\noindent where $W_{\mathrm{vol}}$ and $W_{\mathrm{iso}}$ are the volumetric and isochoric isotropic contributions, respectively. Additionally, $W_{\mathrm{aniso}}$ is the anisotropic strain energy density contribution from the fibers. The isotropic contributions $W_{\mathrm{vol}}$ and $W_{\mathrm{iso}}$ are given by a compressible Neo-Hookean model \cite{simo_2006_book}
\begin{equation}
\begin{split}
    W_{\mathrm{vol}} &= \frac{\kappa}{2} \left(\frac{1}{2}\left(J^2 - 1\right) - \ln{J}\right), \\ 
    W_{\mathrm{iso}} &= \frac{\mu}{2}\left(J^{-\sfrac{2}{d}}\tr\left({\bm{C}}\right) - d \right),
\end{split}
\label{eqn:w_iso}    
\end{equation}
\noindent where $\kappa$ and $\mu$ are the bulk and shear moduli of the solid, $J$ is the Jacobian determinant of the deformation gradient $\bm{F} = \bm{I} + \nabla_{\bm{X}} \bm{u}$, $\bm{I}$ is the identity tensor, $\tr\left(\cdot \right)$ is the trace operator, $\bm{C} = \bm{F}^T\bm{F}$ is the right Cauchy-Green deformation tensor and $d$ is the number of spatial dimensions. The Young's modulus and the Poisson's ratio of the matrix are defined from $\mu$ and $\kappa$ as $E = \frac{9 \kappa \mu}{3 \kappa + \mu}$ and $\nu = \frac{3\kappa - 2 \mu}{2 \left(3 \kappa + \mu \right)}$. The anisotropic strain energy density contribution is given by \cite{nolan_2014}
\begin{equation}
    W_{\mathrm{aniso}} = \frac{k_1}{k_2} \left[ \exp \left( k_2 \left(H - 1\right)^2 \right) - 1\right],
    \label{eqn:w_solid}
\end{equation}
\noindent where $H = \tr(\bm{C}\bm{A})$ is an invariant associated with the fibers, $k_1$ is a material parameter with dimensions of stress that defines the strength of anisotropy and $k_2$ is a dimensionless parameter. We assume that the fibers do not contribute to the material's mechanical response under compression. In Eq.~\eqref{eqn:solidmech_linearmom}, $\bm{P}$ is computed from $W$ as $\bm{P} = 2 \bm{F} \frac{\partial W}{\partial \bm{C}}$.

\subsection{Dimensionless governing equations of the fluid-solid interaction problem}

{We solve the multi-component fluid-solid interaction problem by rescaling the units of length, time, and mass with respect to the droplet radius $R$, visco-capillary time-scale $\frac{\eta_1 R}{\gamma_{\mathrm{LA}}}$ and $\rho R^3$, respectively. With this non-dimensionalization, the coupled system described by Eqs.~\eqref{eqn:nsch} and ~\eqref{eqn:w_solidgeneral} can be characterized by eight dimensionless numbers: a) Ohnesorge number of the liquid $Oh = \frac{\eta_1}{\sqrt{\rho \gamma_{\mathrm{LA}} R}}$ which is the ratio of inertio-capillary to inertio-viscous time scale; b) viscosity ratio of liquid to air $\hat{\eta} = \frac{\eta_1}{\eta_2}$; c) Cahn number $Cn = \frac{\epsilon}{R}$ which is the ratio of the diffuse interface length scale to the droplet radius; d) Peclet number $Pe = \frac{R^2}{M \eta_1}$ which is the ratio of advection to diffusion; e) elastocapillary number $\zeta = \frac{\gamma_{\mathrm{LA}}}{E R}$, which quantifies the strength of elastocapillary effects; f) dimensionless anisotropy strength $\Upsilon = \frac{k_1 \eta_1^2}{\rho_0^s \gamma_{\mathrm{LA}}^2}$; g) Poisson's ratio $\nu$; and h) $k_2$.}

\noindent {We use a superscript * to denote dimensionless quantities. For consistency, we also apply the superscript * to quantities that were already dimensionless before non-dimensionalization, such as $c, \bm{F}$ and $\bm{C}$. We neglect the gravity forces and write the dimensionless governing equations for the fluid-solid interaction as,}
\begin{subequations}
    \begin{align}
        &\nabla^*  \cdot \bm{v}^* = 0, 
        \label{eqn:nsch_divfree_nondim} \\
        &\partial_t^* \bm{v}^* + \bm{v}^* \cdot \nabla^* \bm{v}^*  = \nabla^* \cdot \bm{\sigma}^{f*}, 
        \label{eqn:nsch_momeqn_nondim} \\ 
        &\begin{aligned}
            \partial_t^* c^* + \nabla^*\cdot(\bm{v}^* c^*) &= \frac{1}{Pe} \Delta^* \left(-\frac{3}{2} Cn \ \Delta^* c^* \right) \\
            & + \frac{1}{Pe} \Delta^* \left(\frac{24}{Cn} c^*(1 - c^*)(1 - 2c^*) \right),
        \end{aligned}
        \label{eqn:nsch_phfield_nondim}
    \end{align}
    \label{eqn:nsch_nondim}
\end{subequations}
$\!\!\!$where 
\begin{equation}
\bm{\sigma}^{f*} = - p^* \bm{I} + 2 Oh^2 \ \mathcal{G}(c^*)  \ \nabla^{s*} \bm{v}^* - \frac{3}{2} \ Cn \ Oh^2 \ \nabla^* c^* \otimes \nabla^* c^*,
\end{equation}
and $\mathcal{G}(c^*) = c^* + \frac{1}{\hat{\eta}} (1 - c^*) $. The dimensionless linear momentum balance for the solid is
\begin{equation}
    \left. \partial_t^*\partial_t^* \bm{u}^* \right|_{\bm{X}} = \nabla_{\bm{X}}^* \cdot \bm{P}^*,
    \label{eqn:solidmech_linearmom_nondim}
\end{equation}
$\!\!\!$ where 
\begin{equation*}
    \begin{aligned}
        \bm{P}^* &= \frac{Oh^2\left(\sfrac{\rho_1}{\rho_0^s}\right)}{2 \zeta (1+\nu)} \bm{F}^* (J^*)^{-\sfrac{2}{d}}\Bigl(\bm{I} - \frac{1}{d} \tr\left({\bm{C}^*}\right)(\bm{C}^*)^{-1}\Bigr) \\
        & + \frac{Oh^2\left(\sfrac{\rho_1}{\rho_0^s}\right)}{6 \zeta (1+\nu)} \bm{F}^*\left((J^*)^2 - 1 \right) (\bm{C}^*)^{-1} \\
        & + \frac{4 \Upsilon}{k_2} \bm{F}^* \exp \left( k_2 \left(H^* - 1\right)^2 \right) \left(H^* - 1 \right)\bm{A}^*.
    \end{aligned}
\end{equation*}

We solve the governing equations using a body-fitted fluid-structure algorithm. First, we rewrite the fluid equations in Arbitrary Lagrangian-Eulerian form \cite{donea_2004}. We then recast the governing equations in weak form and discretize them using Isogeometric Analysis \cite{hughes_etal_cmame_2005_iga}. We use a space of splines that is conforming at the fluid-solid interface and perform time integration using the generalized-$\alpha$ method \cite{jansen_2000}. {We provide more details of our computational method in the Appendix.}

\section{Results}

\subsection{Fibrotaxis enables gradient-free, spontaneous droplet motion}

\begin{figure*}[ht!]
    \centering
    \includegraphics[width=.9\linewidth]{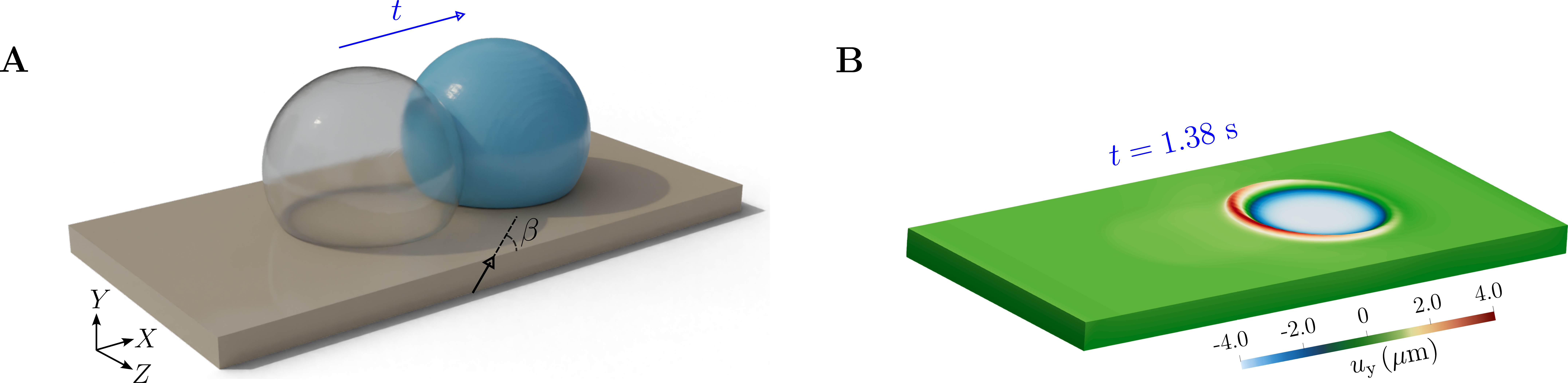}
    \caption{{Three-dimensional simulation of droplet fibrotaxis}. (A) shows the positions of the droplet at different times. The initial position of the droplet is indicated by a semi-transparent spherical cap, while the droplet in blue is at time $t = 1.38$ s. 
    The droplets are represented by isosurfaces of the phase field. The black solid arrow depicts the fibers' orientation. Note that we do not  model the fibers discretely; rather, the model assumes that there are infinite fibers continuously distributed. (B) shows the solid deformation (magnified $3 \times$) colored by the dimensionless vertical solid displacement $u_\mathrm{y}$. We use a non-wetting droplet with $\theta = 120^{\circ}$ and a radius of $160 \ \mu \mathrm{m}$. The fluids properties are $\gamma_{\mathrm{LA}} = 46 \ \mathrm{mN/m}$, $\rho = 1260 \ \mathrm{kg/m^3}$, $\eta_1 = 1.412 \ \mathrm{Pa \cdot s}$ and $\eta_2 = 1.85 \times 10^{-5} \ \mathrm{Pa \cdot s}$. The solid properties are $\rho_0^s = 1000 \ \mathrm{kg/m^3}$, $E = 5$ kPa, $\nu = 0.25$, $k_1 = 40$ kPa, $k_2 = 7$ and $\beta = 60^{\circ}$. The initial solid geometry is a rectangular prism of size $1000 \ \mu \mathrm{m} \times 500 \ \mu \mathrm{m} \times 50 \ \mu \mathrm{m}$. The numerical parameters are $M = 2 \times 10^{-11} \ \mathrm{m^3s/kg}$, $\epsilon = 20 \ \mu \mathrm{m}$ and $\Delta t = 100 \ \mu \mathrm{s}$. The values we choose correspond to the following dimensionless parameters: $Oh = 14.67$, $\hat{\eta} = 7.6 \times 10^4$, $Cn = 0.125$, $Pe = 906.52$, $\zeta = 0.058$ and $\Upsilon = 3.77 \times 10^4$. The initial solid geometry is a rectangular prism of size $1000 \ \mu \mathrm{m} \times 50 \ \mu \mathrm{m} \times 500 \ \mu \mathrm{m}$. We perform computations in a box of size $1000 \ \mu \mathrm{m} \ \times 400 \ \mu \mathrm{m} \ \times 500 \ \mu \mathrm{m}$ that includes the solid and fluid domains.
    }
    \label{fig:droplet_motion_3d}
\end{figure*}

We study fibrotaxis with droplet and solid properties similar to those used for durotaxis \cite{style_etal_2013} and elastocapillary wetting experiments \cite{style_etal_prl_2013}. A 
droplet surrounded by air is placed on a soft solid. Before deformation, the solid geometry is that of a rectangular prism of size $1000 \ \mu \mathrm{m} \times 50 \ \mu \mathrm{m} \times 500 \ \mu \mathrm{m}$; see Fig.~\ref{fig:droplet_motion_3d}A. The selected properties of the droplet and solid yield an elastocapillary length $\sfrac{\gamma_{\mathrm{LA}}}{E}$ equal to $9.2 \ \mu \mathrm{m}$, which is within the elastocapillary wetting range \cite{style_etal_prl_2013}. {The elastic properties of the solid are similar to that of the spin-coated silicone gel used in \cite{style_etal_2013} while the solid's material properties defining the anisotropy strength are similar to those observed in soft biological tissues \cite{holzapfel_etal_ejma_2002}. In our simulation, we use a transversely isotropic material with one family of fibers oriented in the direction $\bm{d} = \cos \beta \ {\bm{e}}_1 + \sin \beta \ {\bm{e}}_2$, where ${\bm{e}}_1$ and ${\bm{e}}_2$ are the unit vectors in the $x-$ and $y-$directions and $\beta$ is the angle of the fibers with respect to $\bm{e}_1$.} The droplet, shown in a semi-transparent color is initially at rest (Fig.~\ref{fig:droplet_motion_3d}A), but the compliance and anisotropy of the solid substrate induce droplet motion immediately after the simulation starts. Fig.~\ref{fig:droplet_motion_3d}B shows that at time $t = 1.38 \ \mathrm{s}$
the droplet has moved parallel to the $x$-axis toward the right-hand side of the solid surface. Fig.~\ref{fig:droplet_motion_3d}B shows a scaled view of the solid's deformed configuration at $t = 1.38 \ \mathrm{s}$ 
with the image colored by vertical solid displacements. We observe a depression of the solid under the droplet, a ridge at the triple line, and negligible displacements everywhere else. Because the solid fibers are parallel to the $xy$\nobreakdash-plane and the droplet is a spherical cap, the solid deformation is symmetric with respect to the $z$\nobreakdash-axis. However, the solid's anisotropy breaks the symmetry with respect to the $x$\nobreakdash-axis resulting in a much larger solid deformation on the left-hand side---this symmetry breaking drives droplet motion. 

\begin{figure*}[ht!]
    \centering
    \includegraphics[width=0.8\linewidth]{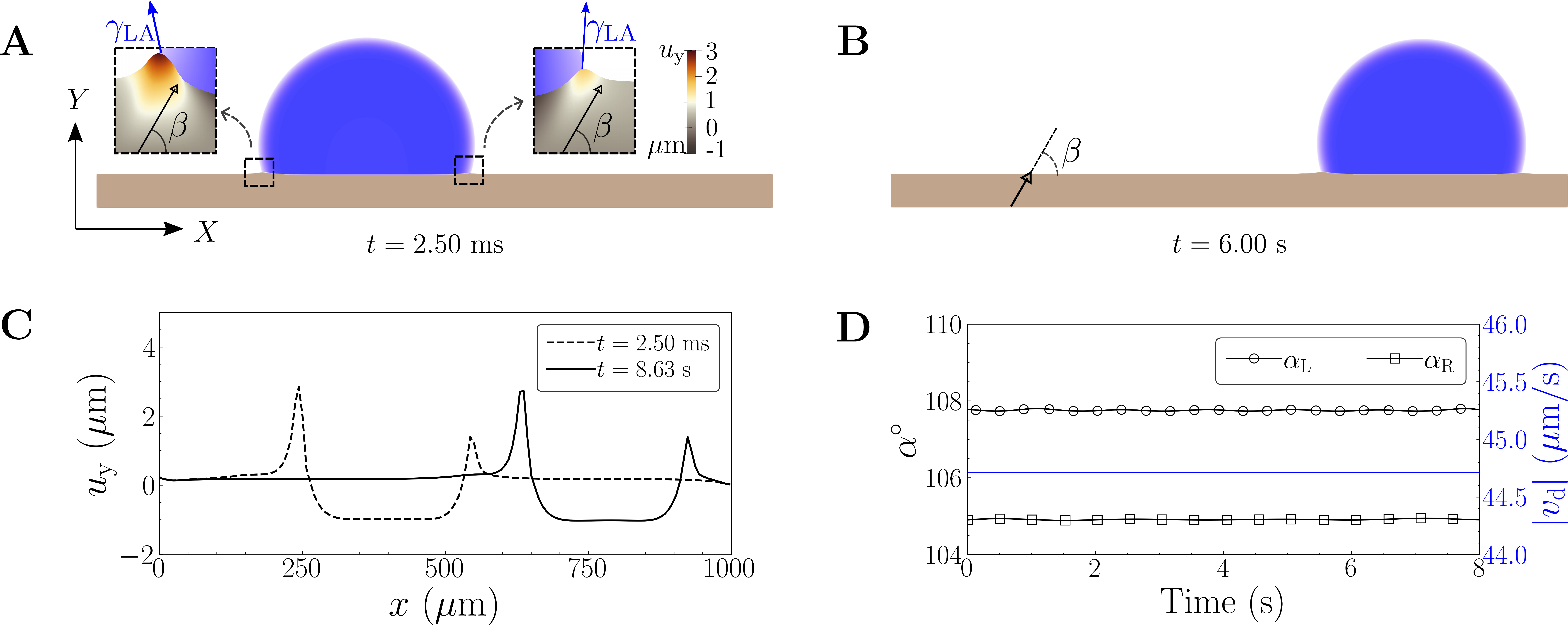}
    \caption{{Mechanism of droplet fibrotaxis} (A-B) show droplet positions at two times. The insets in (A) show the solid deformation ($10 \times$ magnified) and the orientation of the capillary forces acting at the wetting ridges. {The black solid arrows in (A-B) depict the fibers' orientation.} (C) {shows the} vertical displacement of the solid $u_y$ along the fluid-solid interface. (D) shows the time evolution of droplet's velocity ($v_\mathrm{d}$) and the apparent contact angles at the left ($\alpha_{\mathrm{L}}$) and right ($\alpha_{\mathrm{R}}$) contact lines. We plot the droplet velocity using the data of droplet position over time. The variation of apparent contact angles here is a moving average that filters out noise in the data resulting from inaccurate measurements of the contact line due to the diffuse interface. We use a droplet of radius $160 \ \mu \mathrm{m}$ surrounded by air. The droplet is non-wetting with $\theta = 105^{\circ}$. The fluids properties are identical to those used in Fig.~\ref{fig:droplet_motion_3d}. For the solid, we choose $\rho_0^s = 1000 \ \mathrm{kg/m^3}$, $E = 5$ kPa, $\nu = 0.25$, $k_1 = 50$ kPa, $k_2 = 7$,  $\beta = 60^{\circ}$ and a thickness of $50 \ \mu \mathrm{m}$. We also use $M = 2 \times 10^{-11} \ \mathrm{m^3s/kg}$, $\epsilon = 20 \ \mu \mathrm{m}$ and $\Delta t = 25 \ \mu \mathrm{s}$. {The values we choose correspond to $Oh = 14.67$, $\hat{\eta} = 7.6 \times 10^4$, $Cn = 0.125$, $Pe = 906.52$, $\zeta = 0.058$ and $\Upsilon = 4.71 \times 10^4$. We perform our computations in a box of size $1000 \ \mu \mathrm{m} \times 500 \ \mu \mathrm{m}$.}
    }
    \label{fig:mechanism_1}
\end{figure*}

To better understand the mechanism of fibrotaxis, we perform an analogous two-dimensional simulation that permits a more detailed analysis; see Fig.~\ref{fig:mechanism_1}. The droplet, initially at rest on the left-hand side of the solid surface, spontaneously moves to the right. Fig.~\ref{fig:mechanism_1}A shows the system at a very early time---at this point the droplet's motion is negligible, but approximately 6 s later, it has reached the right end of the solid surface (Fig.~\ref{fig:mechanism_1}B). The insets in Fig.~\ref{fig:mechanism_1}A show that the right triple point has a much smaller displacement than the left one. This is a result of capillary forces pulling up the solid more parallel to the fibers on the right triple point, which leads to a stiffer response of the solid. This trend remains throughout the entire simulation as illustrated in Fig.~\ref{fig:mechanism_1}C, which shows the vertical displacement of the solid along the fluid-solid interface. Thus, although our solid is homogeneous, the solid response is stiffer on one side of the droplet and at this point motion occurs due to the same mechanism as in durotaxis; see \cite{style_etal_2013}. However, a critical advantage of fibrotaxis is that it is a gradient-free mechanism, unlike other transport processes that require gradients of stiffness (durotaxis) \cite{style_etal_2013}, temperature, or wettability \cite{daniel_wetgradtemp_sci_2001}. The gradient-free mechanism is advantageous because it is devoid of spatial length scales that would set hard limits on the maximum transport length or maximum velocity. {The fibrotaxis transport length is, in principle, limited only by dissipative mechanisms such as droplet evaporation or contact angle hysteresis, which we have neglected here. Additionally, the gradient-free nature of fibrotaxis allows simple control of the droplet velocity. As shown in Fig.~\ref{fig:mechanism_1}D, the droplet velocity ($v_\mathrm{d}$) remains steady for the length-scale considered in our simulations. Fig.~\ref{fig:mechanism_1}D also shows that the variations in the left and right contact angles ($\alpha_\mathrm{L}$ and $\alpha_\mathrm{R}$) are smaller than $\sim1\%$. We have confirmed from our simulations that the total free energy (which is the analogue of entropy for an isothermal system) of the coupled fluid-solid system is consistently non-increasing over time.}

\begin{figure}
    \centering
    \includegraphics[width=\linewidth]{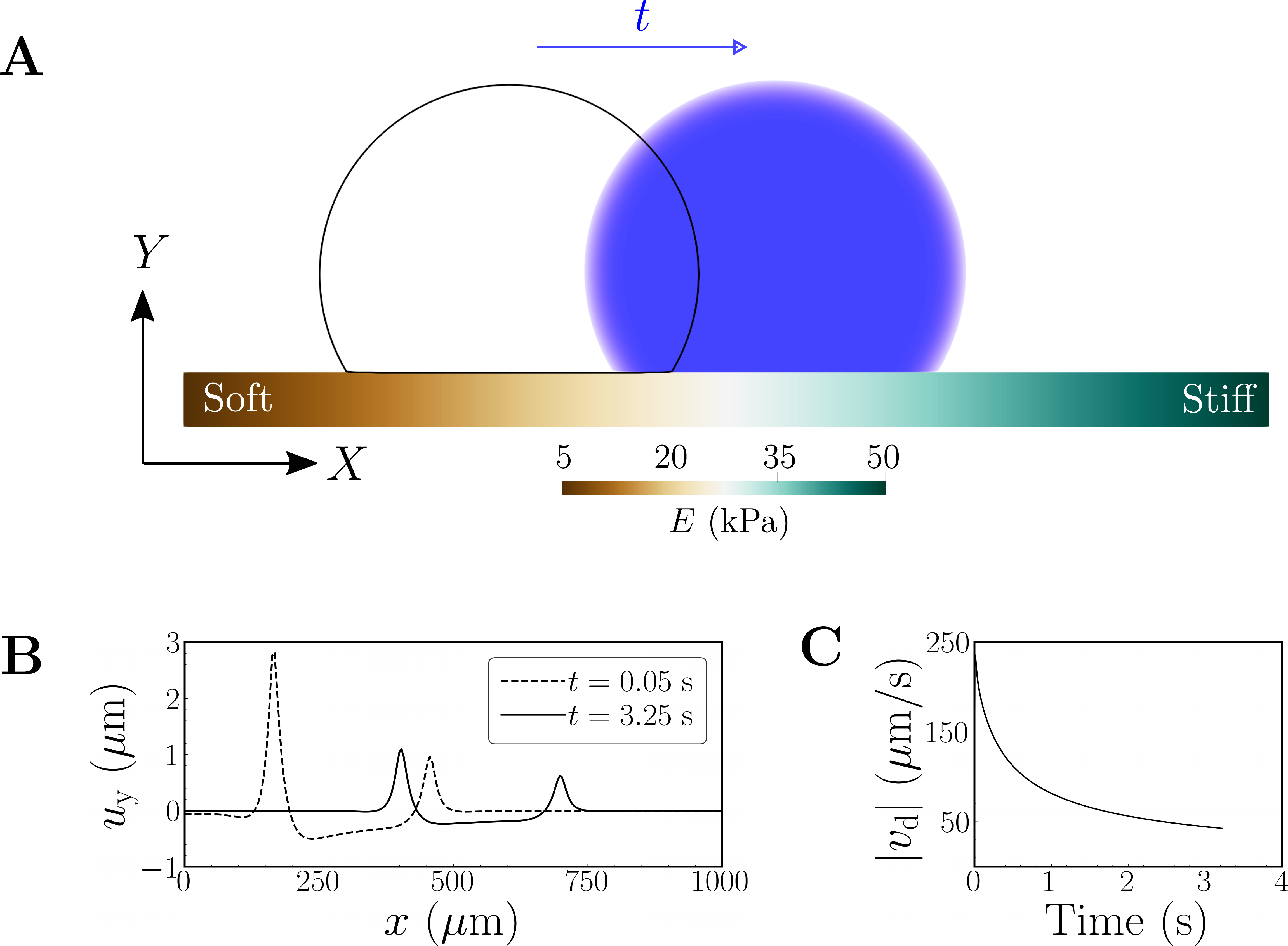}
    \caption{Droplet motion driven by durotaxis. (A) shows the droplet positions at two times. {The initial position of the droplet is indicated by a black spherical cap, while the droplet in blue is at time $t = 3.1$ s.} (B) {shows the} vertical solid displacement $u_\mathrm{y}$ at the fluid-solid interface. (C) {shows the} time evolution of the droplet velocity. We use a non-wetting droplet with $\theta = 120^{\circ}$ and radius of $180 \ \mu \mathrm{m}$, and place it in contact with a heterogeneous solid of thickness $50 \ \mu \mathrm{m}$. The fluids properties are identical to those used in Fig.~\ref{fig:droplet_motion_3d}. We assume the solid is isotropic and its stiffness varies linearly from $E = 5 \ \mathrm{kPa}$ to $E = 50 \ \mathrm{kPa}$. We also use $\nu = 0.25$, $M = 2 \times 10^{-11} \ \mathrm{m^3s/kg}$, $\epsilon = 25 \ \mu \mathrm{m}$ and $\Delta t = 25 \ \mu \mathrm{s}$. {The values we choose correspond to $Oh = 13.82$, $\hat{\eta} = 7.6 \times 10^4$, $Cn = 0.14$, $Pe = 1147.4$, $\zeta \in \left[0.05, 0.005\right]$ and $\Upsilon = 0$. We perform our computations in a box of size $1000 \ \mu \mathrm{m} \times 500 \ \mu \mathrm{m}$.}}
    \label{fig:durotaxis}
\end{figure}

To better illustrate the advantages of a gradient-free mechanism, we perform a durotaxis simulation of a {glycerol} droplet; see Fig.~\ref{fig:durotaxis}. The substrate has a stiffness gradient and is softer on the left-hand side. The droplet moves towards the right end of the solid. The absence of left-to-right symmetry can be seen in Fig.~\ref{fig:durotaxis}B, which shows the vertical solid displacements at the fluid-solid interface at two time instants. As the droplet moves towards the stiffer part, the displacements on both triple points become smaller, leading to a lower driving force that results in a decreasing droplet velocity (Fig.~\ref{fig:durotaxis}C). When the droplet reaches parts of the solid that are very stiff, the problem reverts to that of a droplet on a rigid solid and the droplet motion stops. Therefore, because the solid's stiffness on the left-hand side cannot be made arbitrarily small, the maximum transport distance or the maximum velocity are severely limited by constraints emanating from a gradient-based transport mechanism.

\subsection{The interplay of wettability and fiber orientation regulates droplet fibrotaxis}

\begin{figure}
   \centering
   \includegraphics[width=0.8\linewidth]{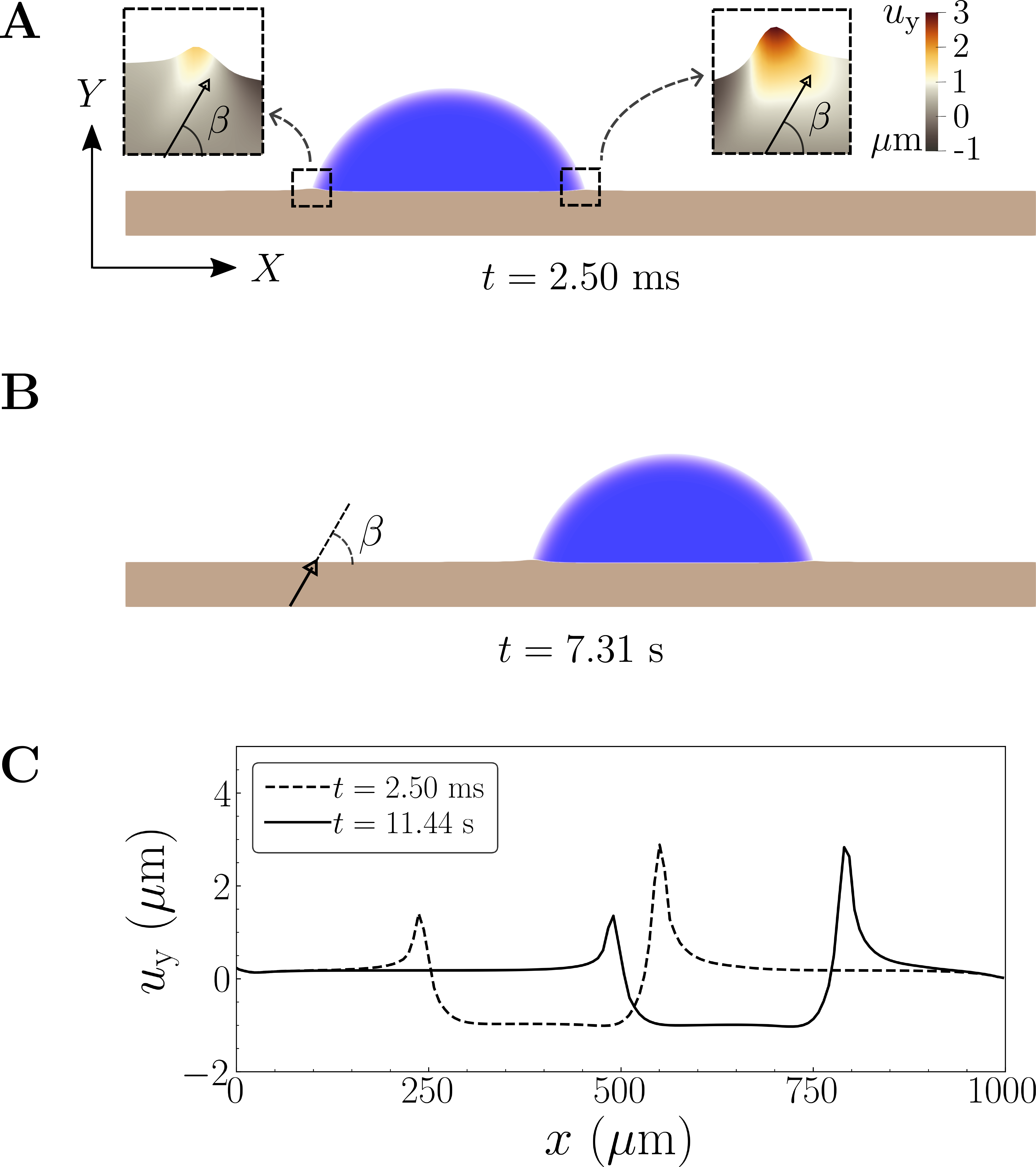}
   \caption{(A-B) Droplet positions at two different times. The insets in (A) show the solid deformation ($10 \times$ magnified) at the wetting ridges. {The black solid arrows in (A-B) depict the fibers' orientation.} (C) {shows the} vertical displacement of the solid $u_y$ along the fluid-solid interface. We use a droplet of radius $160 \ \mu \mathrm{m}$. The droplet is wetting with $\theta = 75^{\circ}$. The properties of fluid, solid, numerical and geometrical parameters are identical to those used in Fig.~\ref{fig:mechanism_1}. {The values we choose correspond to $Oh = 14.67$, $\hat{\eta} = 7.6 \times 10^4$, $Cn = 0.125$, $Pe = 906.52$, $\zeta = 0.058$ and $\Upsilon = 4.71 \times 10^4$.} We perform our computations in a box of size $1000 \ \mu \mathrm{m} \times 500 \ \mu \mathrm{m}$.  
   }
   \label{fig:mechanism_2}
   \vspace{-.5cm}
\end{figure}

The effect of wettability on droplet durotaxis was elucidated in \cite{bueno_2018a}. When wetting droplets are placed on a substrate with a stiffness gradient, they migrate to the soft part. However, non-wetting droplets undergoing durotaxis move to the stiffer part of the substrate. This implies that if we keep the solid's mechanical properties unchanged, motion can be inverted going from a wetting to a non-wetting droplet. In fibrotaxis, wetting (respectively, non-wetting) droplets also move to the part of the substrate with {a} softer (respectively, stiffer) response. However, this implies that if we keep the solid surface unchanged, {the droplet motion cannot} be inverted by changing the fluid's wettability. This is illustrated in Fig.~\ref{fig:mechanism_2}. The two fluids and the mechanical properties of the solid are identical to those in Fig.~\ref{fig:mechanism_1}, but now the solid has more affinity to the fluid, which results in a wetting droplet. The simulation results show motion towards the right-hand side just like in the non-wetting droplet; see Fig.~\ref{fig:mechanism_2}B. However, although the droplet moves towards the right in Figs.~\ref{fig:mechanism_1} and \ref{fig:mechanism_2}, the two processes are different. Indeed, two events are inverted in Fig.~\ref{fig:mechanism_2} with respect to Fig.~\ref{fig:mechanism_1} that have opposing effects and result in invariant motion direction: first, the displacements in the wetting case are larger on the right triple point while in the non-wetting case are larger on the left triple point; second, in the wetting case the droplet moves toward the side with a softer response while in the non-wetting case the droplet moves toward the side with stiffer response. In both the wetting and the non-wetting cases, the motion is in the direction of a lower apparent contact angle.

\subsection{Control of droplet fibrotaxis}

\begin{figure*}[ht!]
    \centering
    \includegraphics[width=0.75\linewidth]{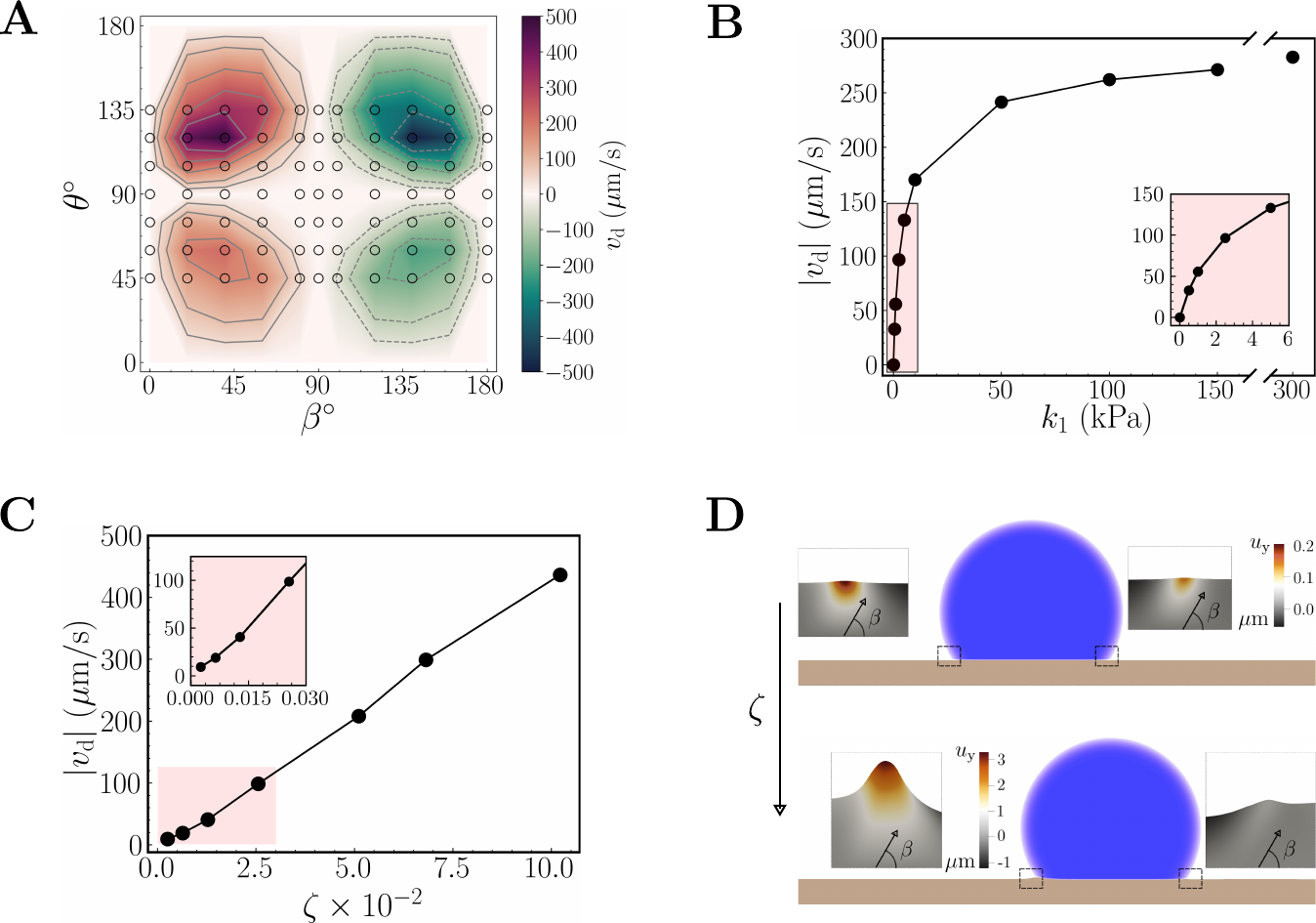}
    \caption{{Control of droplet fibrotaxis. (A) shows the dependence of droplet velocity on fiber angle $\beta$ and wettability $\theta$. Each circle in the plot indicates a numerical simulation for a $\theta-\beta$ pair. We use a droplet with a radius of $180 \ \mu \mathrm{m}$ surrounded by air. The material parameters of the solid are $E = 5$ kPa, $\nu = 0.25$, $k_1 = 50$ kPa and $k_2 = 7$.
    (B) shows the variation of droplet velocity with the strength of anisotropy measured by the parameter $k_1$ (see Eq.~\eqref{eqn:w_solid}). We use a non-wetting droplet with $\theta = 120^{\circ}$ and radius of $160 \ \mu \mathrm{m}$, and place it in contact with air. The material parameters of the solid are $E = 10$ kPa, $\nu = 0.25$, $\beta = 40^\circ$ and $k_2 = 7$.
    (C) shows the variation of droplet velocity with the elastocapillary number $\zeta$. We use $\theta = 120^\circ$ and $\beta = 60^\circ$. The material parameters of the solid are $E \in \left[2.5, 100\right]$ kPa, $\nu = 0.25$, $k_1 = 50$ kPa and $k_2 = 7$. (D) shows the solid deformation ($10 \times$ magnified) at the left and right wetting ridges for two different values of $\zeta$ from (C): $0.068$ and $0.0026$. The black solid arrows in (D) depict the fibers' orientation. For (A-C), we use fluids properties identical to those used in Fig.~\ref{fig:droplet_motion_3d}. Additionally, we use a solid thickness of $50 \ \mu \mathrm{m}$, $M = 2 \times 10^{-11} \  \mathrm{m^3s/kg}$, $\epsilon = 25 \ \mu \mathrm{m}$ and $\Delta t = 25 \ \mu \mathrm{s}$. We perform our computations in a box of size $1000 \ \mu \mathrm{m} \times 500 \ \mu \mathrm{m}$.}
    }
    \label{fig:phase_diagram}
\end{figure*}

The proposed mechanism of droplet fibrotaxis and the roles of fiber orientation and wettability that we identified suggest possible control strategies for droplet motion that include directionality and speed. We illustrate this possibility {by} constructing a droplet velocity diagram on the $\theta-\beta$ phase space. Fig.~\ref{fig:phase_diagram} shows the results of 77 high-fidelity simulations (represented by circles in the plot), each corresponding to a $\theta-\beta$ pair. All simulations were performed using the same value for the droplet radius and the same fluid properties. We positioned the droplet at the center of the domain and assumed that its shape corresponds to that of a circular sector at contact angle $\theta$ with the solid---this would represent a quasi-equilibrium solution for a rigid solid. Our simulations reveal that the magnitude of the droplet velocity is maximum when $\theta\approx\beta$ or $\theta\approx\beta+90^\circ$, which represent cases for which the solid's response is stiffest and softest, respectively, on one of the triple points. The maximum velocity magnitudes are not exactly achieved at $\theta=\beta$ and $\theta=\beta+90^\circ$ because the orientation of the fibers changes with deformation and can only be defined unambiguously in the undeformed configuration. When $\beta=0^\circ,\,90^\circ$ or $180^\circ$, the droplet velocity is negligible because the solid's response is the same on both triple points for any contact angle $\theta$. For the same reason, a neutrally-wetting droplet ($\theta=90^\circ$) shows negligible velocity for any fiber orientation. Fig.~\ref{fig:phase_diagram} also indicates that the velocity magnitude is symmetric with respect to $\beta=90^\circ$, but the motion direction is flipped. Thus, for a fixed $\theta$, the fiber orientations $\beta$ and $180-2\beta$ lead to the same velocity magnitude, but opposite motion directions. Another important conclusion that emerges from Fig.~\ref{fig:phase_diagram} is that, for a given fiber orientation, the direction of droplet motion cannot be reversed changing wettability only. Overall, the data show that fibrotaxis permits to achieve a wide range of droplet velocities with controllable direction in microfluidic applications.

Fig.~\ref{fig:phase_diagram}B shows that the {anisotropy strength $k_1$ (see Eq.~\eqref{eqn:w_solid})} can be used as a {third} independent control parameter for the droplet velocity. The plot indicates that the droplet remains immobile for isotropic solids ($k_1=0$) and moves faster as we increase the strength of anisotropy. At small values of $k_1$, the droplet velocity increases sharply, for example, by about $\sim200\%$ when $k_1$ is increased from $\sim\!1$ to $\sim\!6$ kPa. For large values of $k_1$, the droplet's velocity plateaus at about $\sim\!280\,\mu$m/s. {Although we use fixed values of $\theta$ and $\beta$ in Fig.~\ref{fig:phase_diagram}B, we have verified from our numerical simulations that a similar trend in the variation of droplet velocity with $k_1$ remains across a wide range of $\theta$ and $\beta$ values considered in Fig.~\ref{fig:phase_diagram}A.}

{Fig.~\ref{fig:phase_diagram}C shows that the elastocapillary number $\zeta$ serves as a fourth independent control parameter for the droplet velocity. To plot Fig.~\ref{fig:phase_diagram}C, we vary $E$ and keep $\theta$, $\beta$ and $k_1$ fixed. The plot indicates that the droplet velocity increases as we increase the elastocapillary number. At high values of elastocapillary numbers, the droplet velocity decreases sharply, for example by $\sim46\%$ when the elastocapillary number is lowered from $0.102$ to $0.068$. At very low elastocapillary numbers, the droplet remains nearly stationary. The trend in Fig.~\ref{fig:phase_diagram}C is because the droplet velocity is directly proportional to the magnitude of the relative solid displacement at the left and right triple points, a behavior that has been previously observed in durotaxis \cite{style_etal_2013, bueno_2018a} --- this is illustrated in Fig.~\ref{fig:phase_diagram}D. In Fig.~\ref{fig:phase_diagram}D, at an elastocapillary number of $0.068$, the solid displacement at the left triple point is about $\sim\!1300\%$ more than that on the right triple point, resulting in a high droplet velocity; but, at an elastocapillary number of $0.0026$, the solid displacement at the left triple point is only $\sim\!40\%$ more than that on the right triple point, leading to a low droplet velocity. We can also conclude from Fig.~\ref{fig:phase_diagram}C that when the elastocapillary number is theoretically zero, the droplet will remain stationary because the problem corresponds to that of a droplet on a rigid solid.}

\section*{Conclusions}

The ability to move, mix and process small droplets without external energy supply can open many new opportunities in science, engineering and industry, including, but not limited to, i) new ways to perform multiple or sequential processes in microfluidic and lab-on-a-chip devices, ii) innovative mechanisms for enhanced heat transfer that ensure that fresh coolant is continually in contact with the hot surface, iii) better designs for self-cleaning surfaces that use droplet motion to remove dirt and contaminants, iv) efficient methods to guide condensation and harvest water from air in arid regions, and v) pump-free and smaller devices for biofluid manipulation in medical diagnostics.

This paper proposed fibrotaxis---a simple and robust mechanism for transporting droplets on fiber-reinforced deformable solids. Notably, fibrotaxis is a gradient-free transport mechanism that sets it apart from most droplet motion techniques. Using high-fidelity numerical simulations, we demonstrated that fibrotaxis exhibits three key features essential for efficient droplet transport: controllable droplet velocity and droplet direction, long-distance droplet transport, and spontaneity of droplet motion. {We show that adjusting four control parameters --- fiber angle, wettability, anisotropy strength, and elastocapillary number allows us to control droplet velocity. By constructing a velocity diagram, we additionally show that fiber angle and wettability can control droplet direction.} Our results indicate that fibrotaxis is capable of moving droplets of $\sim$200 $\mu$m radius at speeds of $\sim$200 $\mu$m/s using anisotropic solids that can be fabricated through reinforcement of standard silicone gels with silica \cite{mi_etal_matlet_2018} or polymeric \cite{wang_etal_pnas_2019} fibers. This velocity range is appropriate for most {microfluidics, diagnostics, water harvesting and heat exchange applications,} but higher velocities could be achieved, for example, by coating the solid surface with a thin layer of lubricant or by creating surface textures in the solid \cite{coux_pnas_2020}. These methods reduce viscous dissipation, which facilitates faster droplet motion. Anisotropy is ubiquitous in soft gels. Even those gels which are isotropic prior to deformation can exhibit anisotropy under sufficient deformation due to their nonlinear response or structural changes in the material \cite{vader_plos_2009, baker_ijngeom_1984}.
 
We investigated fibrotaxis in its simplest form, which emerges from the use of a single family of rectilinear fibers embedded in a homogeneous solid matrix. {Our theory, extends to other forms of anisotropy, including multiple fiber families with different orientations and curvilinear fibers (see the Supplementary material\dag \ for more details)}. These forms of anisotropy could facilitate the design of predefined droplet trajectories and velocities. Fibrotaxis can be combined with other forms of droplet control, such as wettability patterns \cite{sun_etal_pnas_2020,ahmet_pnas_2021} or transport mechanisms based on the use of magnetic fields \cite{li_etal_sciadv_2020}. Future efforts could also focus on enhancing the present model by incorporating a nonlinear poro-elastic solid, offering a more realistic depiction of anisotropic gels. \\

{\bf Conflicts of interest}
There are no conflicts to declare. \\

{\bf Data availability}
Data for this article are available at \href{https://doi.org/10.4231/7AXT-KP80}{https://doi.org/10.4231/7AXT-KP80}. \\

{\bf Acknowledgements}
S.R.B. thanks Mario de Lucio for valuable discussions on the solid model. This research was supported by the National Science Foundation (Award no. CBET 2012242). This work uses the Bridges-2 system at the Pittsburgh Supercomputing Center (PSC) through allocation MCH220014 from the Advanced Cyberinfrastructure Coordination Ecosystem: Services $\&$ Support (ACCESS) program, which is supported by National Science Foundation grants 2138259, 2138286, 2138307, 2137603, and 2138296. The opinions, findings, and conclusions, or recommendations expressed are those of the authors and do not necessarily reflect the views of the National Science Foundation.

\section*{Appendix}

\subsection*{Computational method}

Here, we briefly present the key ingredients of our computational method, which has been successfully applied and validated through numerical experiments on various elastocapillary wetting problems \cite{bhopalam_cmame_2022, bueno_2018b, bueno_etal_2017}. {In what follows, we use the dimensional form of the governing equations from Eqs.~\eqref{eqn:nsch} and ~\eqref{eqn:solidmech_linearmom} to derive the variational formulation for the FSI problem.}

\subsubsection*{Arbitrary Lagrangian-Eulerian formulation of fluid mechanics equations}

We solve the governing equations using a body-fitted fluid-structure algorithm similar to that used by the authors in \cite{bhopalam_cmame_2022, bueno_2018b}. The fluid equations (see Eq.~\eqref{eqn:nsch}) are re-written in Arbitrary Lagrangian-Eulerian form \cite{donea_2004}, which are given by, 
\begin{subequations}
    \begin{align}
        &\nabla \cdot \bm{v} = 0, &
        \label{eqn:nsch_ale_divfree} \\
        &\rho \Big(\left.\partial_t\bm{v} \right|_{\widehat{\bm{x}}} + \bm{(v - \widehat{v})} \cdot \nabla \bm{v}\Big) = \nabla \cdot \bm{\sigma}^f + \rho \bm{g}, & \label{eqn:nsch_ale_momeqn_2}  \\ 
        &\begin{aligned}
            \left. \partial_t{c} \right|_{\widehat{\bm{x}}} + \bm{(v - \widehat{v})} \cdot \nabla c &= M\gamma_{\mathrm{LA}} \Delta \left(-\frac{3}{2}\epsilon\Delta c \right) \\ 
            & + M\gamma_{\mathrm{LA}} \Delta \left(\frac{24}{\epsilon} c(1 - c)(1 - 2c) \right)
        \end{aligned}
        \label{eqn:nsch_ale_phfield_2}
    \end{align}
    \label{eqn:nsch_ale}
\end{subequations}
$\!\!\!$where $\bm{\widehat{v}}$ is the fluid mesh velocity, obtained as a mapping between the Eulerian and referential coordinates $\widehat{\bm{x}}$. The subscript $\widehat{\bm{x}}$ in Eqs.~\eqref{eqn:nsch_ale_momeqn_2} and \eqref{eqn:nsch_ale_phfield_2} indicates that the time derivative is taken by holding $\widehat{\bm{x}}$ fixed. For the body-fitted fluid-structure algorithm, we update the fluid mesh by solving successive fictitious linear elasticity problems \cite{wick_2011}. Because density variations do not make a significant difference at this scale, we assumed the fluid density to be constant.

\subsubsection*{Boundary conditions and fluid-solid interface conditions for the FSI problem}

In our numerical simulations, we assume that the initial shape of the droplet corresponds to that of a circular sector at contact angle $\theta$ with the solid---this would represent a quasi-equilibrium solution for a rigid solid. We perform our simulations on prismatic domains. We position the outer boundary of our computational domain far away from the droplet. We impose the following boundary conditions at the lateral fluid boundaries: a) zero normal velocity, b) zero tangential traction and c) zero phase-field flux. However, at the top fluid boundary, we set all fluid velocity components to zero. On the lateral solid boundaries, we impose zero normal displacement and zero tangential traction. But, at the bottom solid boundary, we set all solid displacement components to zero. 

For the coupled fluid-structure interaction problem, we impose the following coupling conditions at the fluid-solid interface: a) wettability condition $\nabla c \cdot \bm{n}^f = |\nabla c|\cos\theta$, b) kinematic condition $\bm{v} = \frac{\partial \bm{u}}{\partial t}$ and c) traction balance $\bm{\sigma^f}\bm{n}^f = \bm{\sigma^s}\bm{n}^f$ where, $\bm{n}^f$ is the unit normal vector at the fluid-solid interface pointing in the direction from fluid to solid, and $\bm{\sigma}^s = J^{-1} \bm{P} \bm{F}^T$ is the solid Cauchy stress tensor. For weak imposition of the wetting boundary condition, we split Eq.~\eqref{eqn:nsch_ale_phfield_2} into two second-order equations by defining the chemical potential $\mu = -\frac{3}{2}\gamma_{\mathrm{LA}}\epsilon\Delta c + \frac{24}{\epsilon} \gamma_{\mathrm{LA}} c(1 - c)(1 - 2c)$.

\subsubsection*{Variational formulation of the FSI problem}

We multiply the governing equations with weight functions and integrate them over the computational domains in the spatial ($\Omega^f_t$) and referential ($\Omega^s_0$) configurations to obtain the weak form. In our FSI problem, we consider the material and referential configurations to be the same. We denote $\mathcal{V}^f, \mathcal{V}^s$ and $\mathcal{V}^m$ to be the trial function spaces for the fluid, solid and mesh motion problems. We also denote $\mathcal{W}^f, \mathcal{W}^s$ and $\mathcal{W}^m$ to be the weight function spaces in a similar fashion. We state the weak form as follows: find $\{p, \bm{v}, c, \mu\} \in \mathcal{V}^f, \bm{u} \in \mathcal{V}^s$ and $\bm{u}^m \in \mathcal{V}^m$ such that $\forall \{w^p, \bm{w}^v, w^c, w^{\mu}\} \in \mathcal{W}^f, \forall \bm{w}^s \in \mathcal{W}^s$ and $\forall \bm{w}^m \in \mathcal{W}^m$, 
\ $B^f\big(\{w^p, \bm{w}^v, w^c, w^{\mu}\}, \{p, \bm{v}, c, \mu\}; \hat{\bm{v}}\big) + B^s\big(\bm{w}^s, \bm{u}\big) + B^m\big(\bm{w}^m, \bm{u}^m\big) = 0$, where
\begin{gather}
\begin{split}
    B^f\big(&\{w^p, \bm{w}^v, w^c, w^{\mu}\}, \{p, \bm{v}, c, \mu\}; \hat{\bm{v}}\big) 
    = 
    \int_{\Omega^f_t} w^p \nabla \cdot \bm{v} \ \strf{d}\Omega      
    \\
    & +
    \int_{\Omega^f_t} \bm{w}^v \cdot \rho \bigg(\left.\partial_t\bm{v}\right|_{\widehat{\bm{x}}} + 
    \bm{(v - \widehat{v})} \cdot \nabla \bm{v}\bigg) \ \strf{d}\Omega     
    \\ 
    & + 
    \int_{\Omega^f_t} \nabla \bm{w}^v : \bm{\sigma}^f \ \strf{d} \Omega +
    \int_{\Omega^f_t} \bm{w}^v \cdot \rho \bm{g} \ \strf{d}\Omega \\
    & + 
    \int_{\Omega^f_t} w^c \left.\partial_t{c}\right|_{\widehat{\bm{x}}} \ \strf{d}\Omega  
    + 
    \int_{\Omega^f_t} w^c (\bm{v} - \hat{\bm{v}}) \cdot \nabla c \ \strf{d}\Omega \\
    & + 
    \int_{\Omega^f_t} M \nabla w^c \cdot \nabla \mu \ \strf{d}\Omega 
    + 
    \int_{\Omega^f_t} w^{\mu} \mu \ \strf{d}\Omega  \\
    &
    - 
    \int_{\Omega^f_t} \frac{3}{2}\epsilon \gamma_{\strf{LA}} \nabla w^5 \cdot \nabla c \ \strf{d}\Omega \\
    &
    - 
    \int_{\Omega^f_t} w^{\mu} \frac{24}{\epsilon} \gamma_{\strf{LA}} c(1 - c)(1 - 2c) \ \strf{d}\Omega \\
    &
    + 
    \int_{\Gamma_t^{sf}} \frac{3}{2}\epsilon \gamma_{\strf{LA}} w^{\mu} \nabla c \cdot \bm{n}^f \ \strf{d}\Gamma, 
\end{split}
\label{eqn:weak_form_fm}
\end{gather}
$\Gamma_t^{sf}$ is the fluid-solid interface in the spatial configuration,
\begin{equation}
    B^s\big(\bm{w}^s, \bm{u}\big) = \int_{\Omega^s_0} \bigg(\bm{w}^s \cdot \rho_0^s \partial_t^2 \left.\bm{u}\right|_{\bm{X}} + \nabla_{\bm{X}} \bm{w}^s : \bm{P} \bigg) \ \strf{d} \Omega, 
    \label{eqn:weak_form_solid}
\end{equation}
\noindent $B^m\big(\bm{w}^m, \bm{u}^m\big)$ is the variational formulation of the fluid mesh motion and $\bm{u}^m$ is the mesh displacement. For numerical implementation of Eq.~\eqref{eqn:weak_form_fm}, we adopt the variational multiscale method (VMS) \cite{hughes_etal_2018}.  

\subsubsection*{Spatial discretization of the FSI problem}

{We spatially discretize the weak form using Isogeometric Analysis (IGA) \cite{hughes_etal_cmame_2005_iga}. We use IGA because of its proven success in computational phase-field modeling and its ability to use basis functions with controllable inter-element continuity. To derive the semi-discrete variational formulation, we substitute the trial and weight function spaces defined earlier with finite-dimensional trial and weight function spaces such that $\mathcal{W}^f_h \subset \mathcal{W}^f, \mathcal{W}^s_h \subset \mathcal{W}^s$, $\mathcal{W}^m_h \subset \mathcal{W}^m$, $\mathcal{V}^f_h \subset \mathcal{V}^f, \mathcal{V}^s_h \subset \mathcal{V}^s$ and $\mathcal{V}^m_h \subset \mathcal{V}^m$. The solution variables and corresponding weight functions are defined as 
\begin{equation}
\begin{aligned}
        & p_h(\bm{x}, t) = \sum_{A \in I_f} p_A(t)N_A(\bm{x}, t), \\
        & w^p_h(\bm{x}, t) = \sum_{A \in I_f} w^p_A(t)N_A(\bm{x}, t), \\
        & \bm{u}_h(\bm{X}, t) = \sum_{A \in I_s} \bm{u}_A(t)\widehat{N}_A(\bm{X}), \\
        & \bm{w}^s_h(\bm{X}) = \sum_{A \in I_s} \bm{w}^s_A\widehat{N}_A(\bm{X}),
\end{aligned}
\end{equation}
where $p_A, w^p_A, \bm{u}_A, \bm{w}^s_A$ are the control variables, $A$ is a control variable index, $I_f$ and $I_s$ are the index sets of the fluid and solid control variables, respectively. Additionally, $\widehat{N}_A$ (and $N_A$) represent spline basis functions defined on referential (and spatial) configuration of the combined fluid and solid. In our simulations, we use quadratic spline basis functions with $\mathcal{C}^1$-inter-element continuity everywhere except along four parametric lines that enclose the solid domain and include the fluid-solid interface.}

\subsubsection*{Time integration and solution strategy}

We perform time-integration of our equations using the generalized-$\alpha$ method \cite{jansen_2000} and solve the coupled FSI problem using a quasi-direct solution strategy \cite{bazilevs_fsirev_2008}. We use the Newton-Raphson method {with a backtracking linesearch} for solving the nonlinear system of equations while we solve the linear solver in each Newton iteration using Generalized Minimal Residual method (GMRES) with a Jacobi preconditioner. {We define a residual vector for each solution variable by substituting the discrete solutions into the variational formulation (see Eqs.~\eqref{eqn:weak_form_fm} and ~\eqref{eqn:weak_form_solid}). We monitor the convergence of the nonlinear solver by ensuring that at least one of the following conditions is met: a) the relative tolerance of each residual vector is smaller than $10^{-3}$, b) the absolute tolerance of each residual vector is smaller than $5 \times 10^{-6}$ and c) the norm of the change in the solution between each Newton step is smaller than $10^{-3}$. Also, we set the linear iterative solver to converge until the relative reduction in the preconditioned residual norm reaches $10^{-5}$}. We develop our code using the open-source frameworks, PetIGA \cite{PETiga_CMAME} and PETSc \cite{petsc-web-page}.

\subsection*{Validation of computational model}

{Here, we present two validation examples performed using our FSI model and computational method. The first example demonstrates the capability of our model to quantitatively capture the elastocapillary FSI phenomenon. The second example shows that our model can quantitatively reproduce the contact line velocity observed in capillary tube experiments.}

\subsubsection*{Validation of FSI model}

{We verify that our FSI code and formulation produce quantitatively accurate results by simulating the static wetting experiment of a glycerol droplet on a thin sheet of silicone gel \cite{style_etal_prl_2013}. By setting $k_1 = 0$ in our model (see Eq.~\eqref{eqn:w_solid}), we perform numerical simulations in a two-dimensional box such that the length of the box is twice the width. We position the outer boundary of our computational domain far from the droplet.}

{In our simulations, we initially position the droplet at the center of the domain such that the droplet-air interface intersects the undeformed solid. We assume that the initial shape of the droplet is a circular sector that makes a contact angle of $96.24^\circ$ with the solid at equilibrium. We impose zero tangential traction and zero phase-field flux along the boundaries of our domain. At the lateral boundaries of the domain, we impose zero velocity in the normal direction. At the bottom boundary, we impose zero normal displacements for the solid. At the top boundary, we impose zero velocity along both the normal and tangential directions.}

\begin{figure}
    \centering
    \includegraphics[width=0.8\linewidth]{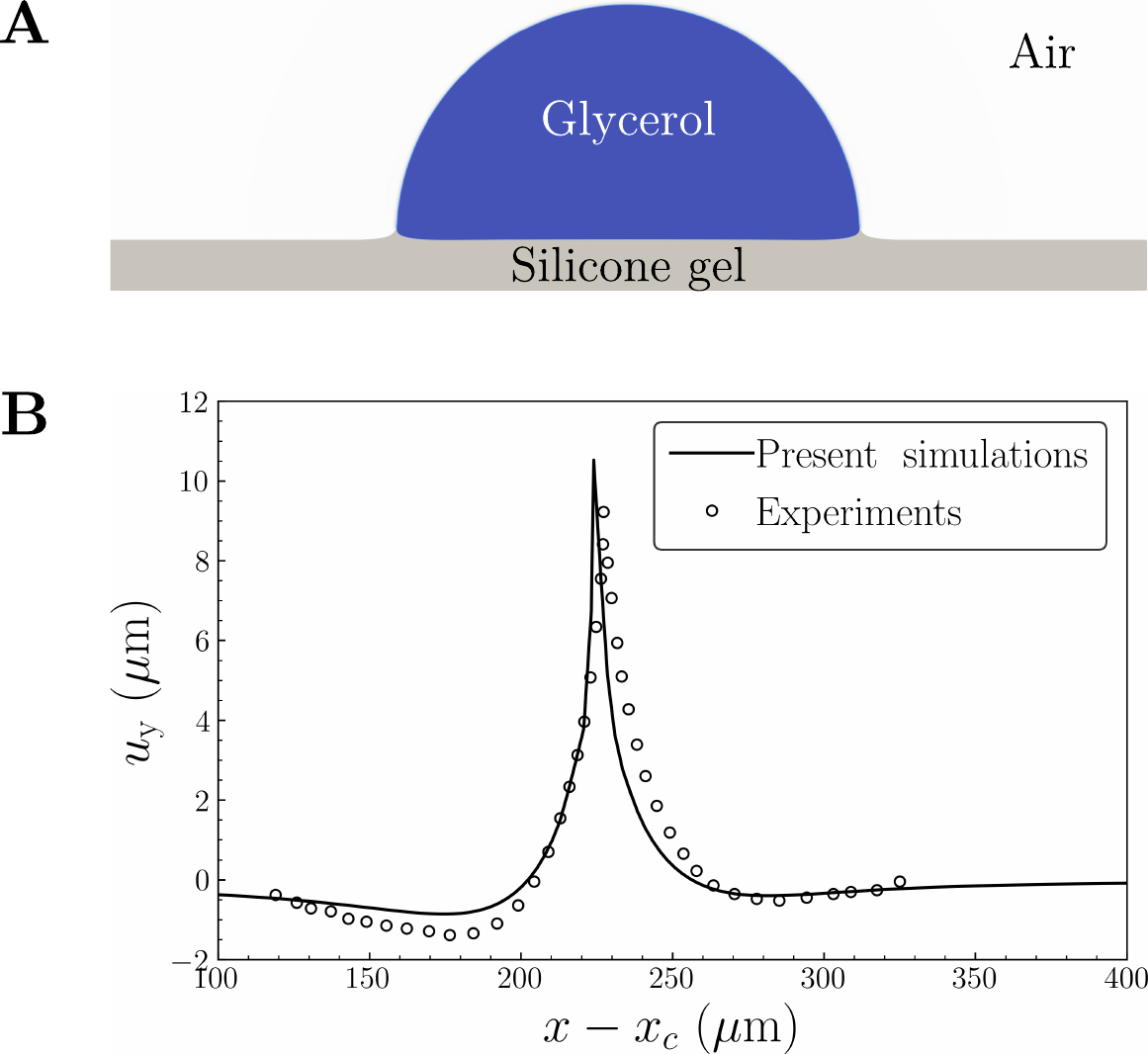}
    \caption{{Static wetting of glycerol droplet on silicone gel. (A) shows how the droplet deforms the solid at steady state. (B) shows the vertical displacement of the solid along the horizontal direction at $t = 300 \ \mu$s. The fluids properties are $\gamma_{\mathrm{LA}} = 46 \ \mathrm{mN/m}$, $\rho = 1260 \ \mathrm{kg/m^3}$, $\eta_1 = 1.4 \ \mathrm{Pa \cdot s}$,  $\eta_2 = 1.4 \mathrm{Pa \cdot s}$. The material properties of the solid are $E = 3.0$ kPa, $\nu = 0.49$ and $\rho_0^s = 1260 \ \mathrm{kg/m^3}$. The numerical parameters are $M = 5 \times 10^{-8} \ \mathrm{m^3s/kg}$,  and $\epsilon = 3 \ \mu \mathrm{m}$. We perform computations in a two-dimensional box of size $1000 \ \mu \mathrm{m} \ \times 500 \ \mu \mathrm{m}$ and spatially discretize it with a mesh of $400 \ \times \ 200$ $\mathcal{C}^1$-quadratic elements.}}
    \label{fig:validation_fsi}
\end{figure}

{Fig.~\ref{fig:validation_fsi} shows our simulation results at $t = 300 \ \mu$s (assumed to be the steady state). Fig.~\ref{fig:validation_fsi}A shows the formation of two wetting ridges at the fluid-solid interface. In Fig.~\ref{fig:validation_fsi}B, we plot the vertical displacement of the fluid-solid interface along the horizontal direction measured from the center of the droplet $x_c$. From Fig.~\ref{fig:validation_fsi}B, it is evident that the data from our simulations are in excellent agreement with experiments \cite{style_etal_prl_2013}.}

\subsubsection*{Validation of contact line dynamics}

{To validate the contact line dynamics captured by our computational method, we simulate the steady-state displacement of glycerol in a rigid capillary tube when air is injected from one side of the tube \cite{zhao_prl_2018}. Since this example serves as a validation case for the fluid mechanics formulation alone, we do not solve the equations of solid dynamics and mesh motion.} 

\begin{figure}
    \centering
    \includegraphics[width=0.8\linewidth]{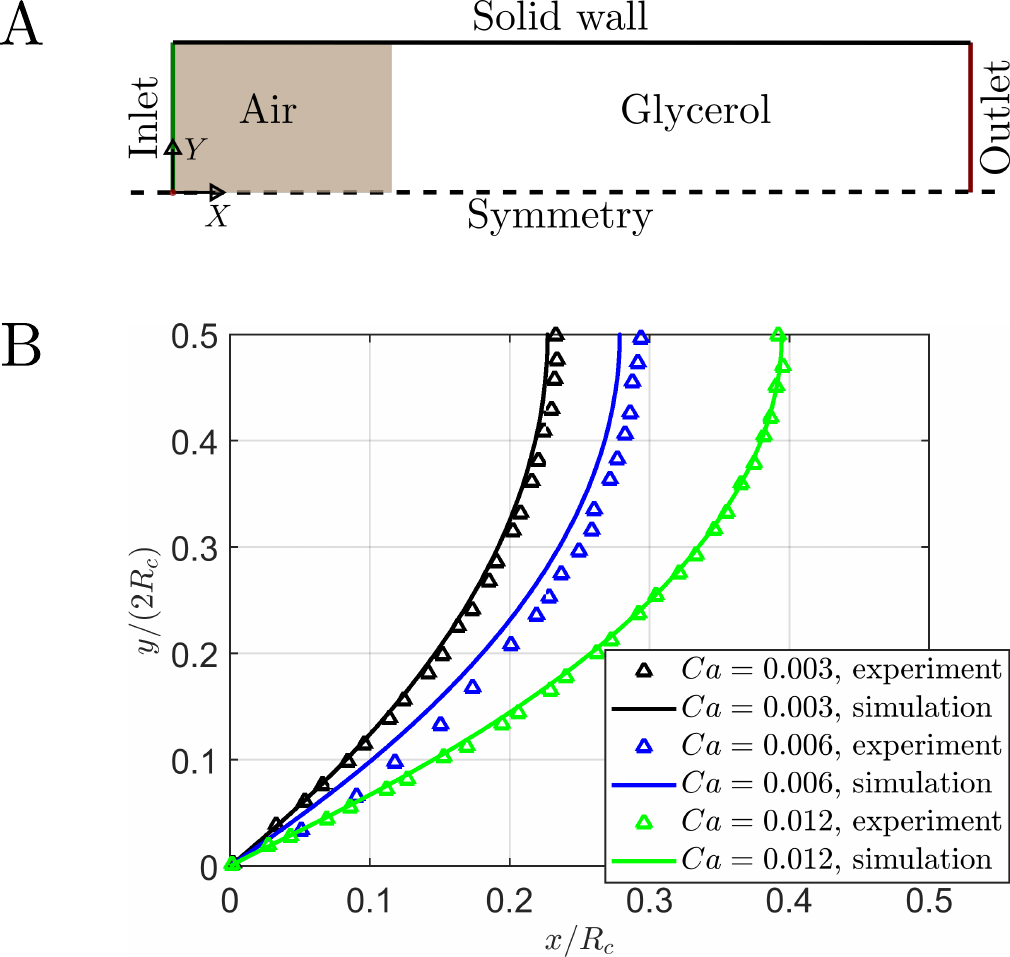}
    \caption{Displacement of glycerol by air in a capillary tube. (A) shows the computational domain. (B) shows the glycerol-air interface profile for three different $Ca$ obtained using experiments \cite{zhao_prl_2018} and present simulations. The radius of capillary tube $R_c$ is $375 \mu$m.
    The fluids properties are $\gamma_{\mathrm{LA}} = 65 \ \mathrm{mN/m}$, $\rho = 1260 \ \mathrm{kg/m^3}$, $\eta_1 = 1.4 \ \mathrm{Pa \cdot s}$,  $\eta_2 = 1.85 \times 10^{-5} \ \mathrm{Pa \cdot s}$ and $\vartheta = 68^\circ$. The numerical parameters are $M = 5 \times 10^{-9} \ \mathrm{m^3s/kg}$, $D_w = 3.327 \ \mathrm{kg/m/s}$ and $\epsilon = 45 \ \mu \mathrm{m}$. We perform computations in a two-dimensional box of size $1875 \ \mu \mathrm{m} \ \times 375 \ \mu \mathrm{m}$ and spatially discretize it with a mesh of $96 \ \times \ 480$ $\mathcal{C}^1$-quadratic elements.}
    \label{fig:validation_rigidtube}
\end{figure}

{Fig.~\ref{fig:validation_rigidtube}A shows the computational setup of our capillary tube. We solve the governing equations given by Eq.~\eqref{eqn:nsch} (or Eq.~\eqref{eqn:nsch_nondim} for the dimensionless governing equations) in axi-symmetric cylindrical coordinates. We perform numerical simulations in a two-dimensional box (see Fig.~\ref{fig:validation_rigidtube}A) such that the length of the box is five times larger than the width of the box. In our simulations, we initially assume that the shape of the air-glycerol interface is flat. We inject air into the inlet boundary of the capillary tube such that the axial velocity is given by $v_{\text{in}} = 2\overline{V}\left (1 - \frac{y^2}{R_c^2}\right)$, where $R_c$ is the radius of the capillary tube and $\overline{V}$ is the average velocity of air at the inlet boundary. At the inlet boundary of our computational domain, we additionally impose a zero radial velocity. At the outlet boundary of our domain, we specify zero radial velocity, zero tangential traction, and zero phase-field flux. At the symmetric boundary, we impose the following symmetry boundary conditions: zero radial velocity, zero tangential traction, and zero phase-field flux. At the solid wall, we impose a no-penetration condition for the velocity, i.e., $\bm{v} = 0$ and a dynamic wettability condition $\nabla c \cdot \bm{n}^f = -D_w \big(\partial_t{c} + \nabla\cdot(\bm{v} c) \big) + \frac{4}{\epsilon} c (1 - c) \cos\vartheta$, where $D_w$ is the dynamic wall mobility that accounts for solid wall relaxation and bulk diffusion near the solid wall while $\vartheta$ is the contact angle made by the glycerol-air interface with the solid.}

{Fig.~\ref{fig:validation_rigidtube}B shows the present simulation results of the air-glycerol interface profile in the capillary tube for different $Ca = \frac{\overline{V} \eta_1}{\gamma_{\mathrm{LA}}}$, which are in excellent agreement with the experiments. It is evident from Fig.~\ref{fig:validation_rigidtube}B that as $Ca$ increases, the air-glycerol interface deforms significantly and bends closer to the symmetry axis. For the range of $Ca$ captured in Fig.~\ref{fig:validation_rigidtube}B, experiments \cite{zhao_prl_2018} indicate that $Ca_{\text{cl}} = Ca$ and $Ca_{\text{tip}} = Ca$, where $Ca_{\text{cl}}$ and $Ca_{\text{tip}}$ are computed from the temporal data of contact line positions at the solid wall and symmetry axis respectively (see Fig.~\ref{fig:validation_rigidtube}A). Table~\ref{tbl:validation_rigidtube} shows the data for $Ca_{\text{cl}}$ and $Ca_{\text{tip}}$ from our simulations. The deviation of the values of $Ca_{\text{cl}}$ and $Ca_{\text{tip}}$ between simulations and experiments is less than $2\%$.}

\begin{table}[ht!]
\small
  \caption{\ Data of $Ca_\text{cl}$ and $Ca_\text{tip}$ predicted from  simulations of air-glycerol displacement in a capillary tube}
  \label{tbl:validation_rigidtube}
  \begin{tabular*}{0.48\textwidth}{@{\extracolsep{\fill}}lll}
    \hline
    $Ca$ & $Ca_\text{cl}$ & $Ca_\text{tip}$ \\
    \hline
    $3.00 \times 10^{-3}$ & $3.01 \times 10^{-3}$ & $3.07 \times 10^{-3}$\\
    $6.00 \times 10^{-3}$ & $6.02 \times 10^{-3}$ & $6.04 \times 10^{-3}$ \\
    $1.2 \times 10^{-2}$ & $1.19 \times 10^{-2}$ & $1.21 \times 10^{-2}$ \\
    \hline
  \end{tabular*}
\end{table}

\bibliographystyle{apsrev4-2}
\bibliography{manuscript}

\end{document}